# Automatic Region Identification over the MMS Orbit by Partitioning n-T space

D. da Silva[1,2,3], A. Barrie[2,4], J. Shuster[2,6], C. Schiff[2], R. Attie[2,5], D. J. Gershman[2], B. Giles[2]

1. Johns Hopkins University, Baltimore, MD, USA
2. NASA Goddard Spaceflight Center, Greenbelt, MD, USA
3. Trident Vantage Systems, Greenbelt, MD, USA
4. Aurora Engineering, Potomac, MD, USA
5. George Mason University, Fairfax, VA, USA
6. University of Maryland, College Park, MD, USA

**Key Points**
1. Automatic identification of regions in the MMS Orbit can be solved with 99.9% accuracy by geometrically partitioning number density vs. ion temperature space by region
2. The best-fit boundaries of these partitions can be elegantly solved for using a machine learning technique called the Support Vector Machine
3. This method provides a fast, simple solution that can be implemented in any language without any machine learning software packages or specialized GPU hardware

**Abstract**
Space plasma data analysis and mission operations are aided by the categorization of plasma data between different regions of the magnetosphere and identification of the boundary regions between them. Without computerized automation this means sorting large amounts of data to hand-pick regions. Using hand-labeled data created to support calibration of the Fast Plasma Instrument, this task was automated for the MMS mission with 99.9% accuracy. The method partitions the number density and ion temperature plane into sub-planes for each region, fitting boundaries between the sub-planes using a machine learning technique known as the support vector machine. This method presented in this paper is novel because it offers both statistical automation power and interpretability that yields scientific insight into how the task is performed.

# Introduction

The Magnetospheric Multiscale Mission (MMS) consists of four spacecraft flying in a tetrahedron formation in a highly elliptical Earth orbit to investigate the process of magnetic reconnection in Earth's space environment (Burch et al, 2016). The orbits of the MMS spacecraft vary throughout the mission to highlight processes in both the dayside (magnetopause) and nightside (magnetotail) regions (Fuselier et al, 2016). The dayside MMS orbit collects data throughout the interior magnetosphere, magnetosheath, and solar wind regions, while the nightside orbit collects data throughout the magnetotail.

Two of the mission's central instruments form the high-time-resolution electron and ion spectrometer suite summarized as the Fast Plasma Investigation (FPI) (Pollock et al, 2016). The Dual Electron Spectrometer (DES) and Dual Ion Spectrometer (DIS) collect full-sky particle velocity distribution functions (VDF) and provide plasma moments such as density, bulk velocity, temperature, and pressure for each respective species.

The method was applied to a magnetopause crossing shown in Figure 1. This magnetopause crossing taken from MMS1 at 2015-09-02 at 13:00 shows how the method performs at a boundary crossing. When applied on a point-by-point basis, the prediction is stable when the region is consistent, and alternates between predictions when the region is changing.

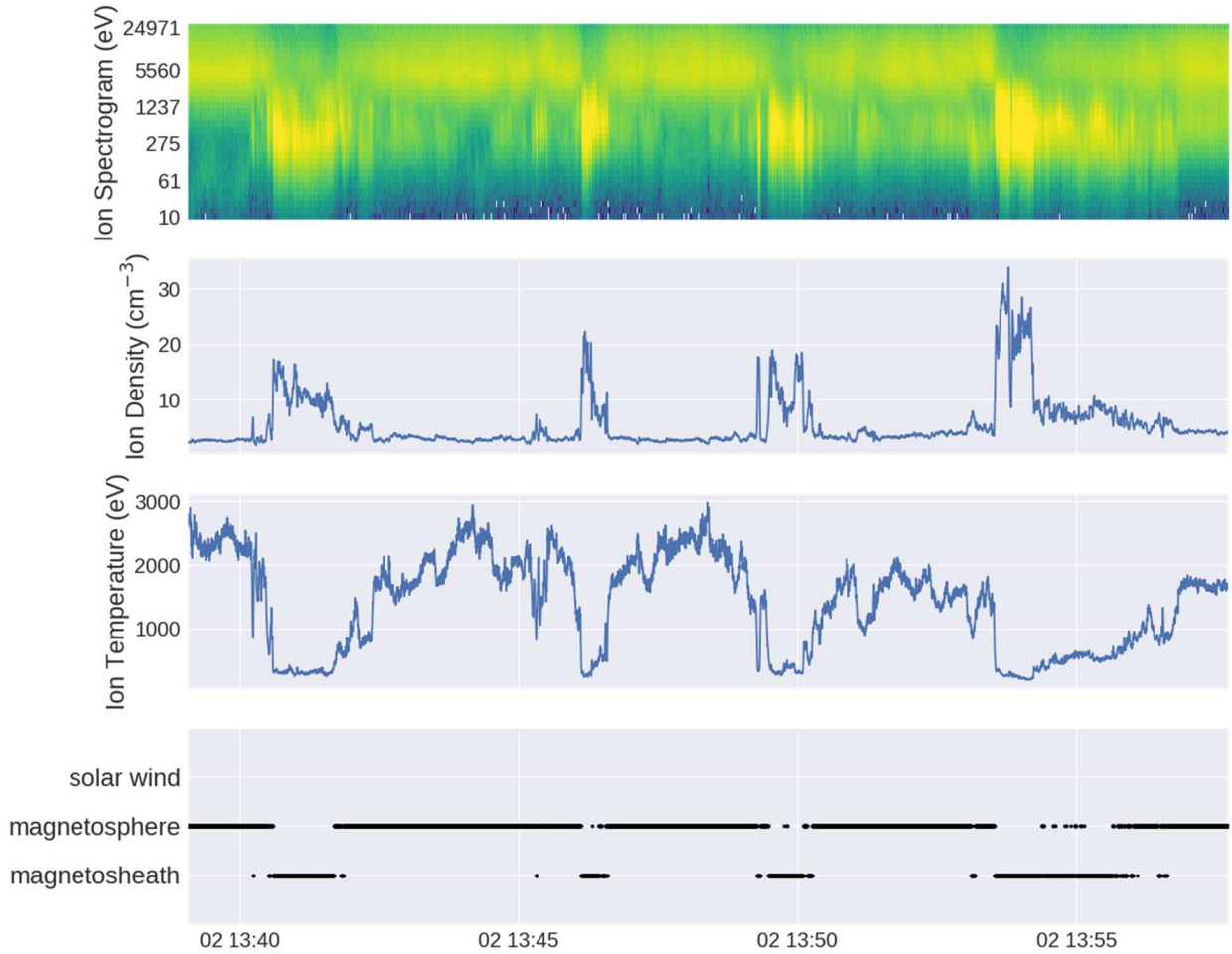

*Figure 1 - Application of method to a magnetopause crossing. In this application the method was applied on a per-point basis. When the method is applied on the border, it alternates between predictions until the prediction is consistent. MMS observed this magnetopause crossing on 2015-09-02 at 13:00 UTC.*

During the beginning of the mission, data was handpicked calibration by a human, who manually reviewed a week of data at a time and sorted each interval into one of several region categories. Over time, examples accumulated into a dataset which associated data with a labeled region. The total dataset of labeled intervals was plotted in the ecliptic plane and colored by the human-given label given in Figure 2.

We automated routine data classification between regions through the method outlined in this paper. Further, we supplied a user interface integrated with the automation. This reduced the time required to keep the instrument calibrated. In this paper, we further discuss the automatic identification in the interest of re-use.

Compared to other techniques considered such as decision trees and convolutional neural networks (Hasties et al, 2005), the method provides a strong argument for both interpretation and computational execution speed. Furthermore, the technique is simple to implement in a language such as IDL or Python and does not require any specialized machine learning software packages or graphics processing unit (GPU) hardware.

Previous studies on this task with THEMIS (Angelopoulos et al, 2009) and ACE (Stone et al, 1998) analyzed a large dataset of the upstream/downstream ratios of the magnetic field and density between the two spacecraft (Jalinek et al, 2012). With these ratios, boundaries between these regions were derived through analysis of the concentrations of samples in the space spanned by these two variables. In contrast, we use a human labeled dataset under the frame of a supervised learning problem. By doing this, we complement the previous work and move in the direction of simplifying the model to use inputs from a single instrument and spacecraft. This simplification provides greater capacity for researchers to reason about the data's nuances, easier integration into an operational ground data system, and grants the option for use in an on-board processor for flight autonomy.

The *Training Data* section will explain how we obtained the data we used to train (optimize) our method. These examples were identified by a human, Dr. C. Schiff, between 2016 and 2019. The *Algorithm* section will give an overview of the method and references to more detailed discussion. The *Evaluation* section will show our justification into the reliability of this method. The *Recommendations* section will discussion recommended use cases for data mining and mission operations, and will also provide guidance on using the algorithm. The *Conclusion* section will summarize the method and suggest next steps to the scientist interested in extending this method.

## Training Data

Between 2016 and 2019, a human scientist labeled data labeled a subset of data in the MMS orbit by its region in order to facilitate instrument calibration using data from a specific region. The three regions labeled were magnetosphere (defined as interior of magnetosphere), solar wind (defined as exterior to the bow shock), and the magnetosheath (defined as between the magnetopause and the bow shock).

In total, there are 18,832 labeled samples positioned around the space environment (Figure 2). These 18,832 samples are individual time step samples originating from 403 start/stop segments. The human labeler labeled each start/stop segment at a time, and then each sample within that segment inherited the label of its parent segment. Each human selection spans an interval of 10-30 seconds, an artifact of how they were originally made to support an instrument calibration algorithm that requires no more than this amount of data.

There are more samples in the magnetosphere than there are in the solar wind or magnetosheath, a phenomenon known in machine learning literature as class imbalance This occurs in our data because more magnetospheric data was needed for our original use case, and therefore the human labeler labeled more of it. The overabundance for one region did not bias the results, and the method does not over-predict magnetosphere.

Of these samples, 90% were saved for used in training (optimizing) the model. 10% were saved for validation of the model. These 10% were not used to tune the model and were used exclusively for testing as they are as independent verification data.

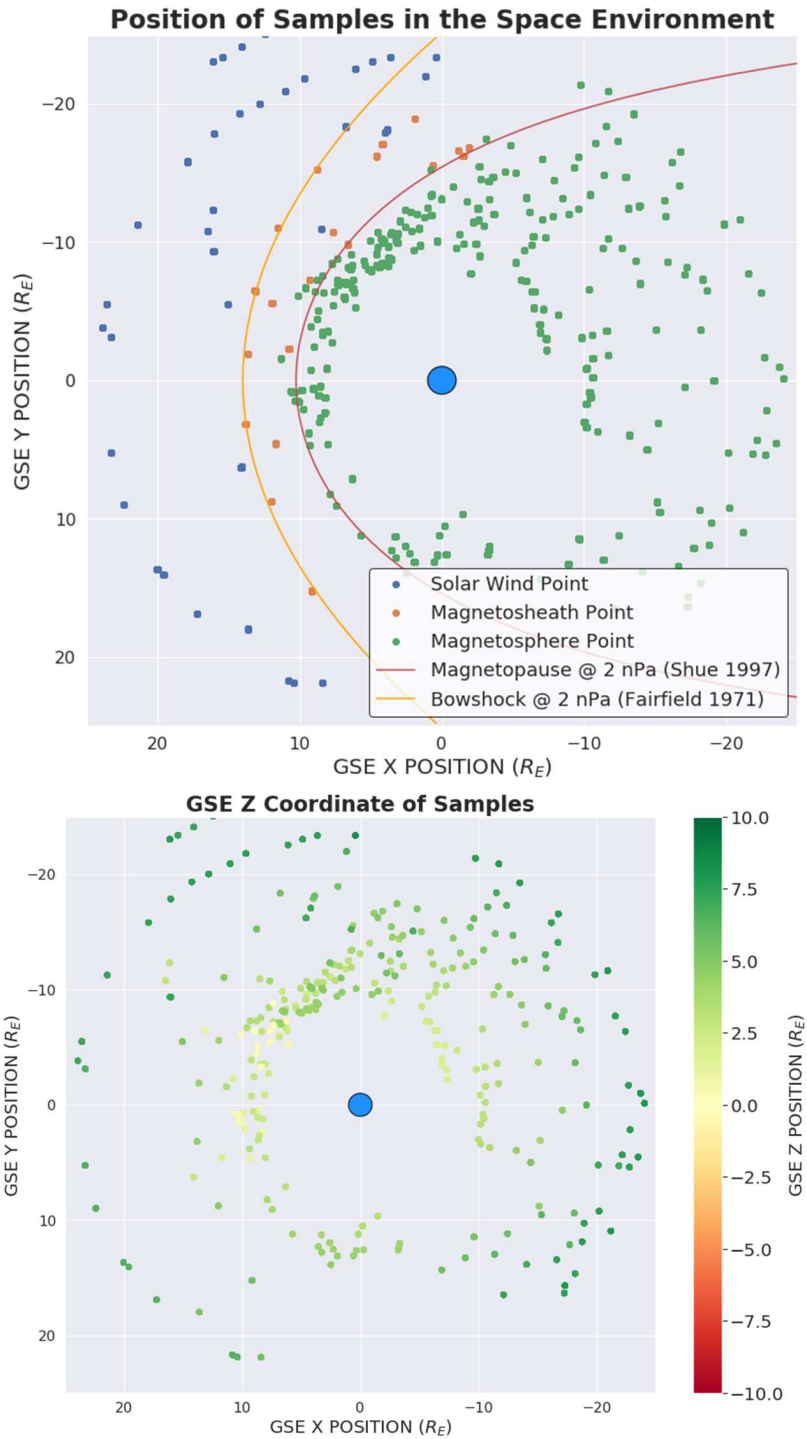

*Figure 2 - Position of Samples in the Space Environment. It is important to note that many samples overlap in space at this zoom level (there are 18,832 samples total, but the above shows 403 dots due to overlap). This plot displays the equatorial magnetopause at 2 nPa solar wind pressure (Shue et al, 1997) and the equatorial bow shock at 2 nPa solar wind pressure (Fairfield et al, 1971). The human-labeled selections use for training data were originally made for instrument calibration purposes, which required the selection to be only 10-30 seconds.*

# Algorithm

Analysis of the ion and electron VDF moments for these methods shows a strong partitioning of the data in the density-temperature space when analyzing both electron and ion data. However, the partitioning was strongest when the ion temperature is used (see supplemental section Figure A.1).

Previous studies on the relationship between density and temperature moments in the magnetosphere have existed for quite some time (Lockwood et al, 1997). A variable they call the *magnetopause parameter* has been used to track the curve through density-temperature space as a spacecraft flies through the magnetopause. Studies of the shape of this curve found a consistent shape of the path traced for all crossings. The ability of these two moments to separate plasmas in the magnetosheath and magnetosphere has a strong intuitive meaning dating back to fundamental physics of gas and plasma state, as encoded in the ideal gas law.

Parameterizing plasma data using the moments n and T is an intuitive way to distinguish and characterize the plasma's state. Our method partitions the number density and ion temperature plane into sub-planes, labeling all samples in each sub-plane as one region ( Figure 3).  The borders between these planes are lines in the log number density and log ion temperature space.

The equation for applying this model can be solved through geometric analysis of these three lines. The simplest application of this technique is to use the lines to calculate a score for each of the three regions, and then assign the identified region to be the region with largest score. This can be done with the following equation:

$$\begin{pmatrix} Magnetosheath\ Score \\ Magnetosphere\ Score \\ Solar\ Wind\ Score \end{pmatrix} = \begin{pmatrix} 4.323 & 1.871 \\ -2.208 & 1.908 \\ -1.702 & -3.866 \end{pmatrix} \begin{pmatrix} Log_{10}(n) \\ Log_{10}(T_i) \end{pmatrix} + \begin{pmatrix} -8.794 \\ -3.636 \\ 9.752 \end{pmatrix} \quad (1)$$

Note that that in the above equation, *n* represents the number density (in units of cm$^{-3}$) and T$_i$ represents the ion temperature (in units of eV).

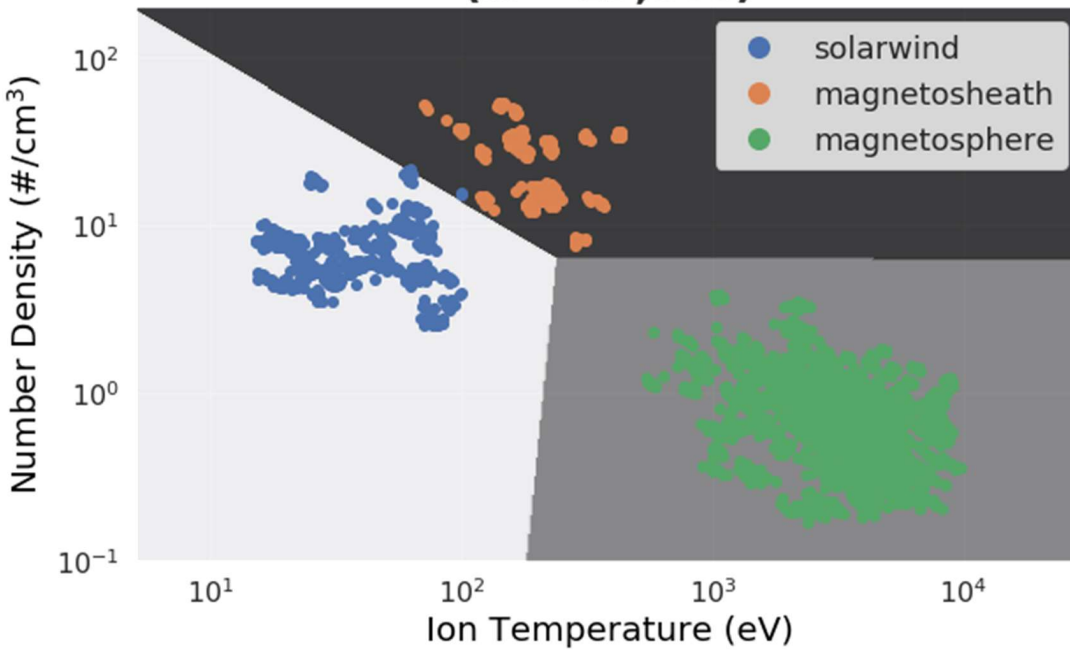

*Figure 3 - Partitioning of Ion Number Density vs Ion Temperature Plane into Regions, with Training Data overlaid. The boundaries are linear in Log-Density and Log-Temperature.*

**Fitting of the Boundaries**
The solution for the boundaries is solved using a statistical technique called the Linear Support Vector Machine. These boundaries represent the solution to a best-fit that separates the data into distinct categories.
The linear support vector machine method appeared in the 1990's and has since had a resurgence in popularity due to elevated attention to machine automation and artificial intelligence. The linear variant of the support vector machine uses straight lines to partition between the categories, however more flexible extensions of the method exist that allow for both curved and non-parametric boundaries. Readers interested in the details of the support vector machine optimization problem, and the variants that exist, are referred to the chapter on Support Vector Machines in Hasties et al, 2005.

**Comparison to Other Methods**
This method has advantages over other numerical and machine learning classification techniques such as decision trees, multi-layer perceptron neural networks (MLP's), and convolutional neural networks (CNN's).

As shown in equation (1), this method can be implemented for a simple point using a single matrix-vector multiplication and vector-vector addition, followed by picking the maximum score. This can be easily implemented in a language such as IDL and Python and has a total of 9 parameters. In this sense, it is very similar to existing methods such as the magnetopause model from Shue 1997 or the bow shock model from Fairfield 1971.

Decision trees are a method which introduces a hierarchy of less-than/greater-than decisions. . While this is simple to understand, the mechanics of implementing it are more complex and many levels are required to capture the diagonal boundary between solar wind and the magnetosheath.

Neural network approaches to solve this problem include the multi-layer perceptron approach operating on the density and temperature moments or the underlying velocity distribution function. The computational performance of a multi-layer perceptron would approach a similar execution speed as this algorithm but would introduce a more difficult model to interpret and would require more parameters to be managed.

A convolutional neural network would also be more difficult to interpret, require even more parameters, and would be a challenge to implement without a machine learning package.

In our method, one can diagnose misclassifications by tracing back the calculation to the density and ion temperature moments. With neural networks, the calculation can easily reach hundreds to millions of calculations, making interpretability a larger challenge (Zhang et al, 2018).

# Evaluation

The method was evaluated on human-labeled samples that were outside the set of samples used to train (optimize) the model. For these samples, an accuracy score of 99.9% was achieved. A per-region break down of the misclassifications follows. The diagonal elements represent correct classifications, while the off-diagonal elements represent misclassifications.

| True <br> Predicted | Magnetosphere | Magnetosheath | Solar Wind |
|---|---|---|---|
| **Magnetosphere** | **2871** | 0 | 0 |
| **Magnetosheath** | 0 | **231** | 3 |
| **Solar Wind** | 0 | 0 | **413** |

*Table 1 - Matrix of correct and incorrect predictions. The units of this table are number of samples. Data from this table is taken from the 10% of human-labeled data that was not included in the model training (optimization). The diagonal elements represent correct identifications, while the off-diagonal elements represent mispredictions. The only mis-predictions were mispredicting solar wind as magnetosheath.*

The three misclassifications were where the human labeled the region as solar wind, while the algorithm labeled the region as magnetosheath. As seen in Figure 3, this occurs at the boundary between solar wind and magnetoheath,which is an acceptable mistake to make.

# Recommendations

This application has use cases for the data mining and automation in mission operations. In this section, comments on extension to other missions and handling mispredictions are also made.

**Data Mining: Boundary Region Searching**

During boundary crossings, the prediction alternates between regions until a stable region is entered. Users of the algorithm may search for boundary crossings by searching for periods of high variability in the region prediction, or search for stable regions by searching for periods of high consistency in the region prediction.

**Mission Operations: Calibration Data from Region**
Instrument calibration often requires certain types of data from different regions to calibrate the instrument. For instance, velocity distribution functions originating in the magnetosphere with strong pitch angle dependency can be used to equate the instrument response in different look-angle directions (Wüest et al, 2007). A hands-off automated system that searches for data in the magnetosphere would be capable of performing this task using our method.

**Extension to Other Missions**
The input parameters in this algorithm are number density in $cm^{-3}$ and ion temperature in eV. In principle, another mission and instrument flying in a similar orbit and measuring these same variables could re-use the model to make the same identification. However, if the orbit includes a region significant to the problem at hand that is not included in this method, care should be taken to integrate it into the final algorithm. The user of the model may wish to find a way to extend the current model or find a way to identify the new region first before passing on the decision to the method presented in this paper. One option would be to manually draw a line separating the new region and its density-temperature space neighbors based on domain knowledge from an expert.

**Interpretation: Handling of Mis-predictions**
A user who is interested in understanding the origin of a misprediction can trace the calculation of the score back to the individual density-temperature moments. From here, a user can judge how the density and temperature contributed to the score. Users can also compare the score between the two highest regions as a metric for a null-decision in close calls.

# Conclusion
Automation of region identification allows scientists to code a task to be done at higher levels of speed possible from a human labeler. With this speed comes potential for larger amounts of identification and machinery to create per-region datasets for subsequent analysis.

One should keep in mind that the applicability this model is restricted to samples in the MMS orbit; which are samples in the ecliptic no further out than around 25 $R_E$. Because MMS explores just the magnetosphere, solar wind, and magnetosheath, one may ask how the method would be extended to include other notable plasma regions such as the ionosphere and Van Allen Radiation Belts. Sufficiently labeled data from the Van Allen Probes (Spence et al, 2013) would be useful to answer this question.

The MMS mission uses a scientist-in-the-loop system for selecting all of its downlinked data (Baker et al, 2016). In this system a scientist reviews low-resolution data and selects a subset of high-interest data that should be downlinked at high-resolution. A lesser known part of this system is that every human selection requires a text comment describing why it was selected. As of 2019, over 48,000 intervals have been selected and annotated using this system. Though there has been little convention outside of FPI calibration for how these text comments are written and formatted, they still exist and provide a repository of labeled data identifying space plasma physics phenomena.

We would like to thank the FPI instrument operations scientists and engineers at Goddard Spaceflight Center for their help labeling and supporting data for experiment.

## Acknowledgments
MMS/FPI flight data, including the training data used for this publication, is available from the MMS Science Data Center at https://lasp.colorado.edu/mms/sdc/public/.

# Appendix

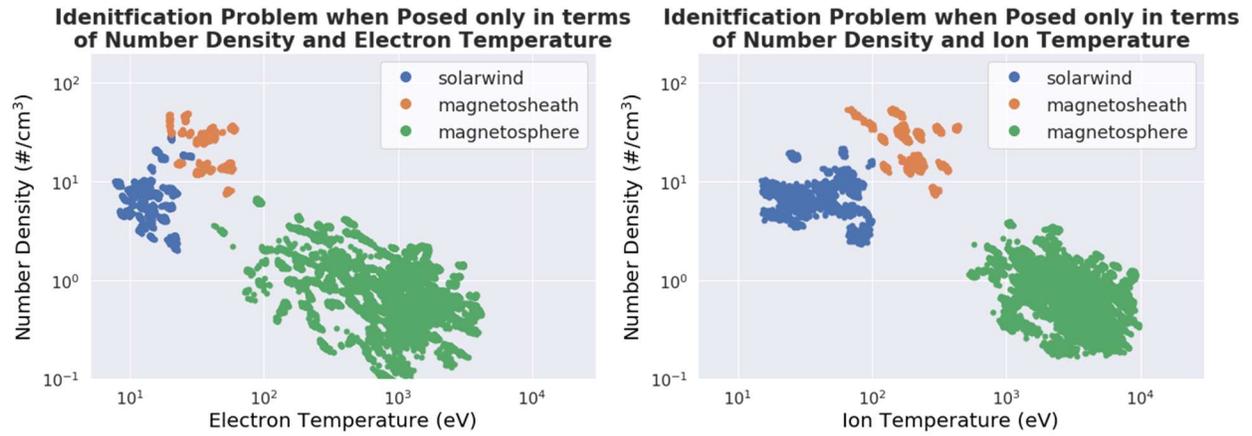

*Figure A.1 - Comparison of how samples cluster in space on the planes of electron number density vs electron temperature and ion number density vs ion temperature. The former ion data had the strongest separation characteristic, while the electron data showed some confusion between the solar wind / magnetosheath regions.*